\documentclass[rmp]{revtex4}
\usepackage{amssymb,amsfonts,amsmath}

\usepackage{hyperref}
\usepackage{graphicx}
\usepackage{amsmath}
\usepackage{natbib}

\newcommand{\EQ}[1]{Eq.~(\ref{eq:#1})}

\newcommand{\FIG}[1]{Fig.~\ref{fig:#1}}

\newcommand{\beast}[1]{\emph{#1}}

\newcommand{\mfit}{\bar{x}}
\newcommand{\pfix}{\phi}

\newcommand{\pop}{n}
\newcommand{\Pfix}{\Phi}
\newcommand{\xs}{\chi}
\newcommand{\ox}{r}
\newcommand{\xo}{\rho}

\newcommand{\mut}{U}

\begin{document}
\title[Draft and selective interference]{Genetic draft, selective
interference, and population genetics of rapid adaptation}
\author{Richard~A.~Neher}
\affiliation{Max Planck Institute for Developmental Biology, T\"ubingen, 72070,
Germany. \\richard.neher@tuebingen.mpg.de}
\date{\today}

\begin{abstract}
To learn about the past from a sample of genomic sequences, one needs to
understand how evolutionary processes shape genetic diversity.
Most population genetic inference is based on frameworks assuming adaptive
evolution is rare. But if positive selection operates on many loci
simultaneously, as has recently been suggested for many species including
animals such as flies, a different approach is necessary.
In this review, I discuss recent progress in characterizing and understanding
evolution in rapidly adapting populations where random associations of mutations
with genetic backgrounds of different fitness, i.e.,~genetic draft, dominate
over genetic drift. As a result, neutral genetic diversity depends weakly on
population size, but strongly on the rate of adaptation or more generally the
variance in fitness. Coalescent processes with multiple mergers, rather than
Kingman's coalescent, are appropriate genealogical models for rapidly
adapting populations with important implications for population genetic
inference.
\end{abstract}
\maketitle 
\tableofcontents

\section{Introduction}
Neutral diffusion or coalescent models
\citep{Kimura:1964p3388,Kingman:1982p28911} predict that genetic diversity at
unconstrained sites is proportional to the (effective) population size $N$ --
for a simple reason:
Two randomly chosen individuals have a common parent with a probability of order
$1/N$ and the first common ancestor of two individuals lived of order $N$
generations ago. Forward in time, this neutral coalescence corresponds to
\emph{genetic drift}. However, the observed correlation between genetic
diversity and population size is rather weak
\citep{Lewontin1974,leffler_revisiting_2012}, implying that processes other than
genetic drift dominate coalescence in large populations. This notion is
reinforced by the observation that pesticide resistance in insects can evolve
independently on multiple genetic backgrounds
\citep{Karasov:2010p35377,labbe_independent_2007} and can involve several
adaptive steps in rapid succession \citep{schmidt_copy_2010}. This high
mutational input suggests that the short-term effective population size of
\beast{D.~melanogaster} is greater than $10^9$ and conventional genetic drift
should be negligible.
Possible forces that accelerate coalescence and reduce diversity are
\emph{purifying} and \emph{positive} selection. Historically, the effects of
purifying selection have received most attention (reviewed by
\citet{Charlesworth:2012p45100}) and my focus here will be on the role of
positive selection.

A selective sweep reduces nearby polymorphims through \emph{hitch-hiking}.
Polymorphisms linked to the sweeping allele are brought to higher
frequency, while others are driven out \citep{Smith:1974p34217}. 
Linked selection not only reduces diversity, but also slows down adaptation in
other regions of the genome -- an effect known as Hill-Robertson interference
\citep{Hill:1966p21029}. Hill-Roberston interference has been intensively
studied in two locus models \citep{Barton:1994p34628} where the effect is quite
intuitive: two linked beneficial mutations arising in different individuals
compete and the probability that both mutations fix increases with the
recombination rate between the loci. Pervasive selection, however, requires
many-locus-models. 
Here, I will review recent progress in understanding how selection at
many loci limits adaptation and shapes genetic diversity. 
Linked selection is most pronounced in asexual organisms.
The theory of asexual evolution is partly motivated by evolution experiments
with microbes, which have provided us with detailed information about the spectrum of
adaptive molecular changes and their dynamics.
I will then turn to facultatively sexual organisms which include many important
human pathogens such as HIV and influenza as well as some plants and nematodes.
Finally, I will discuss obligately sexual organisms, where the effect of linked
selection is dominated by nearby loci on the chromosome.

The common aspect of all these models is the source of stochastic fluctuations:
random associations with backgrounds of different fitness. In contrast to
genetic drift, such associations persist for many generations, which amplifies
their effect. In analogy to genetic drift, the fluctuations in allele
frequencies through linked selection have been termed \emph{genetic draft}
\citep{Gillespie:2000p28513}.
The (census) population size determines how readily adaptive mutations and
combinations thereof are discovered but has little influence on
coalescent properties and genetic diversity. Instead, selection
determines genetic diversity and sets the time scale of coalescence. The
latter should not be rebranded as $N_e$ as this suggests that a rescaled neutral
model is an accurate description of reality. In fact, many features are
qualitatively different. Negligible drift does not imply that selection is
efficient and only beneficial mutations matter. On the contrary, deleterious
mutations can reach high frequency through linkage to favorable backgrounds and
the dynamics of genotype frequencies in the population remains very stochastic.
Genealogies of samples from populations governed by draft do not follow the
standard binary coalescent process. Instead coalescent processes allowing for
multiple mergers seem to be appropriate approximations which capture the large 
and anomalous fluctuations associated with selection.
Those coalescent models thus form the basis for a \emph{population genetics of rapid adaptation} and serve as null-models to
analyze data when Kingman's coalescent is inappropriate. To illustrate
clonal interference, draft, and genealogies in presence
of selection, this review is accompanied by a collection of scripts
based on FFPopSim \citep{zanini_ffpopsim:_2012} 
at
\href{http://webdav.tuebingen.mpg.de/interference}{webdav.tuebingen.mpg.de/interference}.

\section{Adaptation of large and diverse asexual populations}
Evolution experiments (reviewed in
\citet{kawecki_experimental_2012,burke_how_2012}) have demonstrated that 
adaptive evolution is ubiquitous among microbes.
Experiments with RNA viruses have shown that the rate of
adaptation increases only slowly with the population size
\citep{Miralles:1999p47914,deVisser_diminishing_1999}, suggesting that adaptation is limited by
competition between different mutations and not by the
availability of beneficial mutations.
The competition between clones, also known as \emph{clonal interference}, was
directly observed in \beast{E.~coli} populations using fluorescent markers
\citep{Hegreness:2006p11675}. Similar observations have been made in Rich
Lenski's experiments in which \beast{E.~coli} populations were followed for more
that 50000 generations \citep{Barrick:2009p39413}. A
different experiment selecting $>100$ \beast{E.~coli} populations for heat tolerance has
shown that there are 1000s of sites available for adaptive substitutions, that
there is extensive parallelism among lines in the genes and pathways bearing
mutations, and that mutations frequently interact epistatically
\citep{Tenaillon:2012p47907}.
By following the frequencies of microsatellite markers in populations of
\beast{E.~coli}, \citet{Perfeito:2007p34158} estimated the beneficial mutation rate to
be $U_b\approx 10^{-5}$ per genome and generation with average effects of about
$1\%$. Similarly, it has been shown that beneficial mutations are readily
available in yeast and compete with each other in the population for fixation
\citep{Desai:2007p17662,Lang:2011p41792,Kao:2008p47893}.
At any given instant, the population is thus characterized by a large number of
segregating clones giving rise to a broad fitness distribution
\citep{Desai:2007p17662}. The fate of a novel mutation is mainly determined by
the genetic background it arises on \citep{Lang:2011p41792}. Similar rapid
adaptation and competition is observed in the global populations of influenza,
which experience several adaptive substitutions per year
\citep{Bhatt:2011p43255,Strelkowa:2012p48466,Smith:2004p20265}, mainly driven by immune
responses of the host.
In summary, evolution of asexual microbes does not seem to be limited by finding
the necessary single point mutations, but rather by overcoming clonal interference and
combining multiple mutations.

\begin{figure}[htp]
\begin{center}
  \includegraphics[width=0.6\columnwidth]{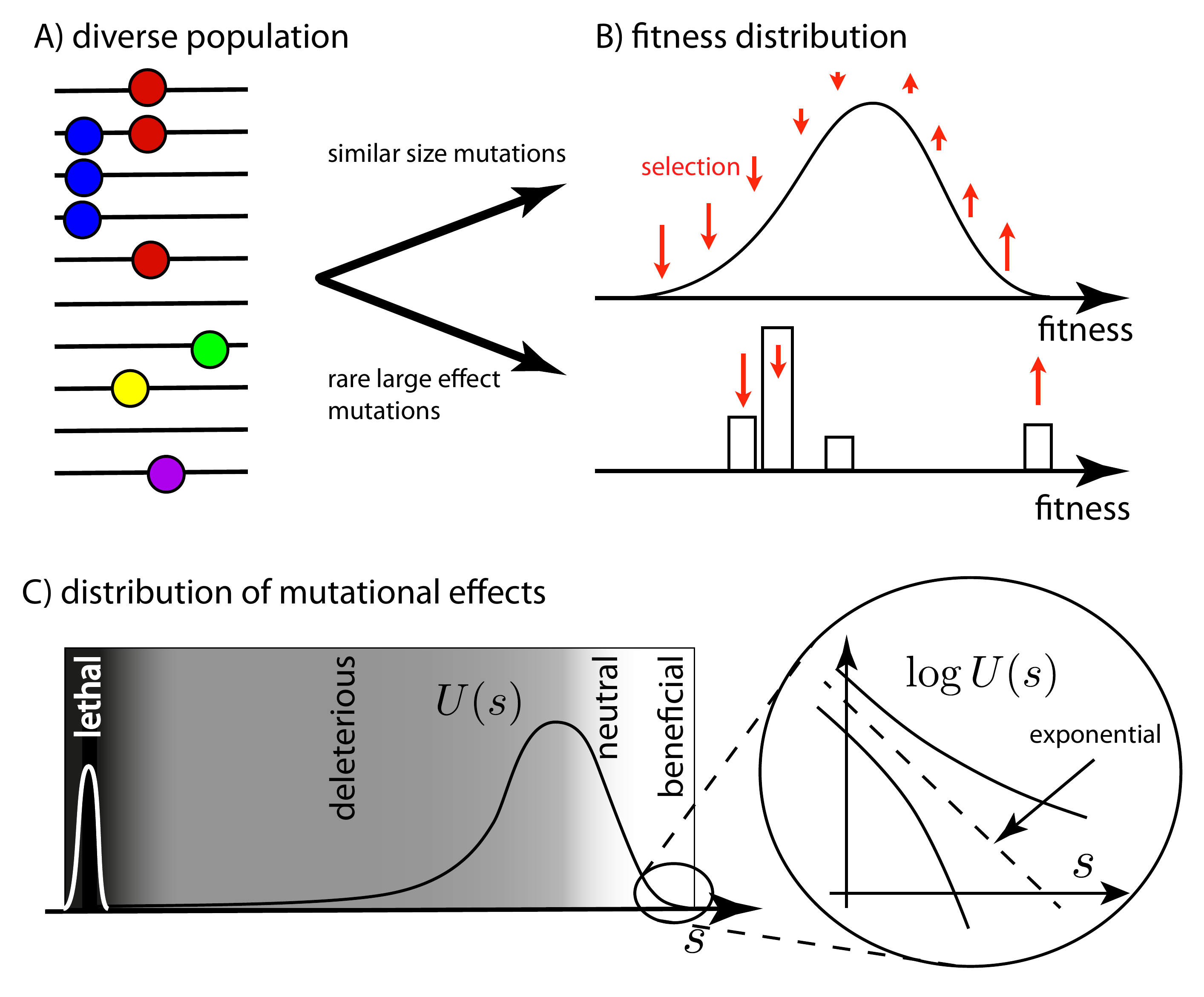}
  \caption[labelInTOC]{Fitness and mutational effect distributions. (a) A
  genetically diverse population will typically harbor variation in fitness. 
  If many mutations have comparable effects on fitness, the resulting
  fitness distribution is smooth and roughly normal (part b, top). If a small
  number of large effect mutation exists, the distribution is multi-modal
  (part b, bottom).
  Mutational effects across the genome are believed to follow a distribution roughly like the one
  sketched in panel (c). A small fraction of mutations are beneficial, the
  majority are neutral or deleterious, and some are lethal. The integral over
  $\mut(s)$ is the total mutation rate $\mut$.  In models
  of adaptive evolution, the high fitness tail of $\mut(s)$, shown into in the
  inset, is the most important part.  If it falls off faster than exponentially,
  the fitness distribution tends to be smooth. Otherwise, the distribution is 
  often dominated by a few large effect mutations. }
  \label{fig:pop_and_distribution_sketch}
\end{center}
\end{figure}

These observations have triggered intense theoretical research on clonal
interference and adaptation in asexuals. In the models studied, rare events,
e.g.~the fittest individual acquiring additional mutations, dramatically affect the future
dynamics. Intuition is a poor guide in such situations and careful mathematical
treatment is warranted. Nevertheless, it is often possible to rationalize the
results in a simple and intuitive way with hindsight, and I will try to present
the important aspects in accessible form.

Our discussion assumes that fitness is a unique function of the genotype.
Thereby, we ignore the possibility of frequency-dependent selection. A diverse
population with many different genotypes can then be summarized by its
distribution along this fitness-axis; see \FIG{pop_and_distribution_sketch}A\&B.
Fitness distributions are shaped by a balance between injection of variation via
mutation and the removal of poorly adapted variants.
Most mutations have detrimental effects on fitness, while only a small minority
of mutations is beneficial. The distribution of mutational effects in RNA virus
has been estimated by mutagenesis \citep{Sanjuan:2004p24501,Lalic:2011p44586}.
Roughly half of random mutations are effectively lethal, while $~4\%$ were found
to be beneficial in this experiment. A distribution of mutational effects,
$\mut(s)$, is sketched in \FIG{pop_and_distribution_sketch}C. General properties
of $\mut(s)$ are largely unknown and will depend on the environment.

\begin{figure}[htp]
\begin{center}
  \includegraphics[width=0.5\columnwidth]{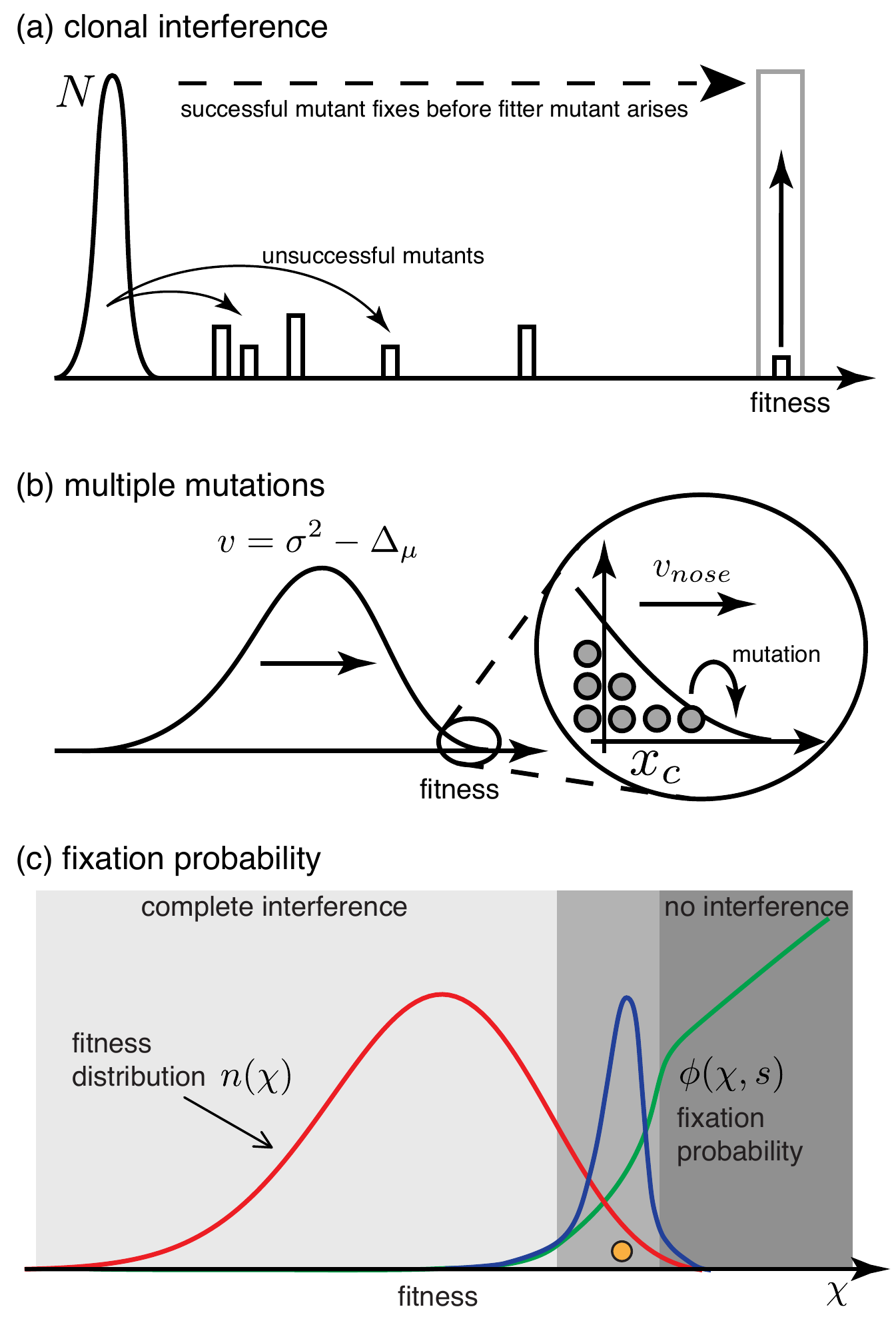}
  \caption[labelInTOC]{Adaptation in asexual populations. (a) If the
  distribution of beneficial mutation has a long tail, the population 
  consists of a small number of large clones and only the mutations with the
  largest effects have a chance of fixing. (b) If many mutations of similar
  effect contribute to fitness diversity, the bulk of the fitness distribution
  can be described by a smooth function that is roughly Gaussian in shape.
  There exists a fittest genotype in the population with no individuals to its
  right. Only mutations close to this high fitness ``nose'' have an appreciable
  chance of fixing. The stochastic dynamics at the nose determines the 
  evolution of the entire population and the speed of the entire population,
  $v$, has to match the speed of the nose, $v_{nose}$, in a quasi-steady state.
  The fixation probability $\pfix(\xs,s)$ of a mutation with effect $s$
  increases with increasing background fitness as sketched in panel (c). A
  mutant in the bulk of the fitness distribution has essentially zero chance of
  taking over the population since many fitter individuals exist.
  In the opposite case when the mutant is the fittest in the population,
  $\pfix(\xs,s)$ is proportional to $\xs+s$ as we would expect in the absence
  of interference. Since there
  are very few individuals with very high fitness, most mutations that fix come from a narrow region (light grey) where the
  product of $\pop(\xs)$ and $\pfix(\xs,s)$, sketched in blue, peaks. Note that
  $\xs$ is Malthusian or log-fitness. Scripts to illustrate interference and
  fixation can be found in the \href{http://webdav.tuebingen.mpg.de/interference/fixation_asex.html}{online
  supplement}.}
  \label{fig:CI_vs_MM}
\end{center}
\end{figure}

Deleterious mutations rarely reach high frequencies but are numerous, while
beneficial mutations are rare but amplified by selection. But in order to spread
and fix, a beneficial mutation has to arise on an already fit genetic background or have a
sufficiently large effect on fitness to get ahead of everybody else.
Two lines of theoretical works have put emphasis either on the large effect
mutations (clonal interference theory) or ``coalitions'' of multiple mutations
of similar effect.
Both approaches, sketched in \FIG{CI_vs_MM} are good approximations depending on
the distribution of fitness effects.

\subsection{Clonal Interference} Consider a homogeneous population in which
mutations with effect on fitness between $s$ and $s+ds$ arise with rate
$\mut(s)ds$ as sketched in \FIG{pop_and_distribution_sketch}C. In a large
population many beneficial mutations arise every generation. In order to fix, a
beneficial mutation has to outcompete all others; see \FIG{CI_vs_MM}A.
In other words, a mutation fixes only if no mutation with a larger effect arises
before it has reached high frequencies in the population. This is the essence of
clonal interference theory by \citet{Gerrish:1998p5933}. The Gerrish-Lenski
theory of clonal interference is an approximation since it ignores the
possibility that two or more mutations with moderate effects combine to
outcompete a large effect mutation -- a process I will discuss below. Its
accuracy depends on the functional form of $\mut(s)$ and the population size
\citep{Park:2007p4335}.
One central prediction of clonal interference is that the rate of adaptation
increases only slowly with the population size $N$ and the beneficial mutation
rate $U_b$.
This is a consequence of the fact that the probability that a particular
mutation is successful decreases with $NU_b$ since there are more mutations
competing. This basic prediction has been confirmed in evolution experiments
with virus
\citep{Miralles:1999p47914,Miralles:2000p47915,deVisser_diminishing_1999}. 
How the rate of adaptation depends on $N$ and $U_b$ is sensitive to the distribution of fitness
effects $\mut(s)$. Generically, one finds that the rate of adaptation is
$\propto (\log NU_b)^\alpha$, where $\alpha$ depends on the properties of
$\mut(s)$ \citep{Park2010}.

Clonal interference theory places all the emphasis on the mutation with the
largest effect and ignores variation in genetic background or equivalently the
possibility that multiple mutations accumulate in one lineage. It is therefore
expected to work if the distribution of effect sizes has a long tail allowing
for mutations of widely different sizes. It fails if most
mutations have similar effects on fitness. A careful discussion of the theory of clonal
interference and its limitations can be found in \citet{Park2010}. 

\subsection{Genetic background and multiple mutations}
If most beneficial mutations have similar effects, a lineage cannot fix by
acquiring a mutation with very large effect but has to accumulate
more beneficial mutations than the competing lineages. 
If population sizes and mutation rates are large enough that many mutations
segregate, the distribution $\pop(x,t)$ of fitness $x$ in the population is
roughly Gaussian, see \FIG{CI_vs_MM}B, and the problem becomes tractable
\citep{Tsimring:1996p19688,Rouzine:2003p33590,Desai:2007p954}. More precisely,
$\pop(x,t)$ is governed by the deterministic equation
\begin{equation}
\label{eq:popdis}
\frac{d}{d t}\pop(x,t) = (x-\mfit)\pop(x,t) + \int \mut(s)[
\pop(x-s,t) - \pop(x,t)]\, ds
\end{equation}
where $(x-\mfit)\pop(x,t)$ accounts for amplification by selection of
individuals fitter than the fitness mean $\mfit$ and elimination of the less fit
ones.
The second term accounts for mutations that move individuals from $x-s$ to $x$
at rate $\mut(s)$. Integrating this
equation over the fitness $x$ yields Fisher's ``Fundamental Theorem of
Natural Selection'', which states that the rate of increase in mean fitness is
\begin{equation}
\label{eq:FT}
\frac{d}{d t}\mfit  = v = \sigma^2 - \Delta_\mu
\end{equation}
where $\sigma^2$ is the variance in fitness and $\Delta_\mu$ is the average
mutation load a genome accumulates in one generation. A steadily moving mean
fitness $\mfit=vt$ suggests a traveling wave solution of the form $\pop(x,t)
=\pop(\xs)$ where $\xs=x-\mfit$ is the fitness relative to the mean.
\EQ{FT} is analogous to the breeder's equation that
links the response to selection to additive variances and co-variances.
In quantitative genetics, the trait variances are determined empirically and
often assumed constant, while we will try to understand how $\sigma^2$ is
determined by a balance between selection and mutation.

To determine the average $v$, we need an additional relation between $v$ and the
mutational input. To this end, it is important to realize that the population is
thinning out at higher and higher fitness and only very few individuals are
expected to be present above some $\xs_c$ as sketched in \FIG{CI_vs_MM}B. The
dynamics of this high fitness ``nose'' is very stochastic and not accurately
described by \EQ{popdis}. However, the nose is the most important part where
most successful mutations arise.
There have been two strategies to account for the stochastic effects and derive an additional relation for the
velocity. (i) The average velocity, $v_{nose}$, of the nose is determined by a
detailed study of the stochastic dynamics of the nose.
At steady state, this velocity has to equal the average velocity of
the mean fitness given by \EQ{FT}, which produces the additional relation required to determine
$v$ \citep{Tsimring:1996p19688,Rouzine:2003p33590,Cohen:2005p45154,Desai:2007p954,Goyal:2012p47382,Brunet:2008p12980}.
(ii) Alternatively, assuming additivity of mutations, $v$ has to equal the
average rate at which fitness increases due to fixed mutations
\citep{Neher:2010p30641,Good:2012p47545} (see \citep{Hallatschek:2011p39697}
for a related idea).
I will largely focus on this latter approach, as it generalizes to sexual
populations below. In essence, we need to calculate the probability of fixation $\Pfix(s,v)$
of mutations with effect size $s$ that arise in random individuals in the
population. $\Pfix$ depends on $v$ and implicitly on the traveling
fitness distribution $\pop(x-vt)$.
Using this notation, we can express $v$ as the sum of effects of mutations
that fix per unit time:
\begin{equation}
\label{eq:selfconsist}
v = \frac{d}{d t}\mfit = N \int \mut(s) \Pfix(s,v) s \,ds
\end{equation}
Note that the mutational input is proportional to the census population size
$N$. To solve \EQ{selfconsist}, we first have to calculate the fixation
probability $\Pfix(s,v)$, which in turn is a weighted average of the fixation
probability, $\pfix(\xs,s)$, given the mutation appears on a genetic background
with relative fitness $\xs$. The latter can be
approximated by branching processes \citep{Neher:2010p30641,Good:2012p47545}.
A detailed derivation of $\pfix(\xs,s)$ is given in the
supplement of \citet{Good:2012p47545}, while the subtleties associated with
approximations are discussed in \citet{fisher_asexual_2013}.
The qualitative features of $\pfix(\xs,s)$ are sketched in \FIG{CI_vs_MM}C.

The product $\pop(\xs)\pfix(\xs,s)$ describes the distribution of
backgrounds on which successful mutations arise. This distribution is often
narrowly peaked right below the high fitness nose (see
\FIG{CI_vs_MM}C). Mutations on backgrounds with lower fitness are
doomed, while there are very few individuals with even higher background
fitness. The larger $s$, the broader this region is.

To determine the rate of adaptation, one has to substitute the results for
$\Pfix(s,v)$ into \EQ{selfconsist} and solve for $v$
\citep{Desai:2007p954,Good:2012p47545}.
A general consequence of the form of the self-consistency condition
\EQ{selfconsist} is that if $\Pfix$ is weakly dependent on $v$, we will find $v$
proportional to $N$. In this case the speed of evolution is proportional to the
mutational input. With increasing fitness variance, $\sigma^2$, the genetic background 
fitness starts to influence fixation probabilities, such that eventually $v$
increases only slowly with $N$. For models in
which beneficial mutations of fixed effect $s$ arise at rate $U_b$, the rate
of adaptation in large populations is given by
\begin{equation}
v \propto \begin{cases}
 s^2 \frac{\log Ns}{(\log U_b/s)^2} & s\gg U_b \\
 (U_bs^2)^{2\over 3}(\log ND^{1/3})^{1/3} & s\ll U_b
\end{cases}
\label{eq:asex_speed}
\end{equation}
\citep{Desai:2007p954,Cohen:2005p45154}. The above has assumed that $s$ is
constant, but these expressions hold for more general models with a
short-tailed distribution $\mut(s)$ with suitably defined effective $U_b$ and
$s$ \citep{Good:2012p47545}.

\paragraph*{Synthesis}
Clonal interference and multiple mutation models both predict diminishing
returns as the population increases,  but the underlying dynamics are rather different.
In the clonal interference picture, population take-overs are driven by single
mutations and the genetic background on which they occur is largely
irrelevant ($\pfix(\xs,s)$ depends little on $\xs$). The mutations that are
successful, however, have the very largest effects. In the multiple mutation
regime, the effect of the mutations is not that crucial, but they have to occur
in  very fit individuals to be successful ($\pfix(\xs,s)$ increases
rapidly with $\xs$). In both models, the speed of adaptation continues to
increase slowly with the population size and there is no hard ``speed limit''.
Distinguishing a speed limit from diminishing returns in experiments is hard
\citep{deVisser_diminishing_1999,Miralles:2000p47915}. 

Whether one or the other picture is more appropriate depends on the distribution of available
mutations $\mut(s)$. If $\mut(s)$ falls off faster than exponential,
adaptation occurs via many small steps
\citep{Desai:2007p954,Good:2012p47545}; if the distribution is
broader, the clonal interference picture is a reasonable approximation
\citep{Park:2007p4335,Park2010}.
The borderline case of an exponential fitness distribution has been investigated
more closely, finding that large effect mutations on a pretty
good background make the dominant contributions
\citep{Good:2012p47545,Schiffels:2011p43799}, i.e., a little bit
of both.

Empirical observations favor this intermediate situation. Influenza evolution
has been analyzed in great detail and is was found that a few rather than
a single mutation drive the fixation of a particular strain
\citep{Strelkowa:2012p48466}. Similarly, evolution experiments suggest that the
genetic background is important, but a moderate number of large effect mutations account for most of
the observed adaptation \citep{Lang:2011p41792}. 

Note the somewhat unintuitive dependence of $v$ on parameters in
\EQ{asex_speed}. Instead of the mutational input $NU_b$ and $s$, $v$ depends on
$Ns$ and $U_b/s$ for $U_b\ll s$. In large populations, the dominant time scale
of population turnover is goverened by selection and is of order $s^{-1}$. $Ns$
and $U_b/s$ measure the strength of reproduction noise (drift) and mutations
relative to $s^{-1}$, respectively (see \citet{Neher:2012p47353} for a
discussion of this issue in the context of deleterious mutations).
In large populations, the infinite sites model starts to break down and the
same mutations can occur independently in several lineages limiting interference
\citep{Bollback:2007p47921,Kim:2005p6895}.

\section{Evolution of facultatively sexual populations}
Competition between beneficial mutations in asexuals results in a slow
(logarithmic) growth of the speed of adaptation with the population size
$N$ (\EQ{asex_speed}). How does gradually increasing the outcrossing rate
alleviate this competition? The associated advantages of sex and recombination have been
studied extensively
\citep{Fisher_1930,Muller_AmericanNaturalist_1932,Crow:1965p1011,Charlesworth:1993p25074,Rice:2001p4303}.
It is instructive to consider facultatively sexual organisms that outcross at
rate $\ox$, and in the event of outcrossing have many independently segregating
loci. Facultatively sexual species are common among RNA viruses, yeasts,
nematodes, and plants.

\begin{figure}[htp]
\begin{center}
  \includegraphics[width=0.7\columnwidth]{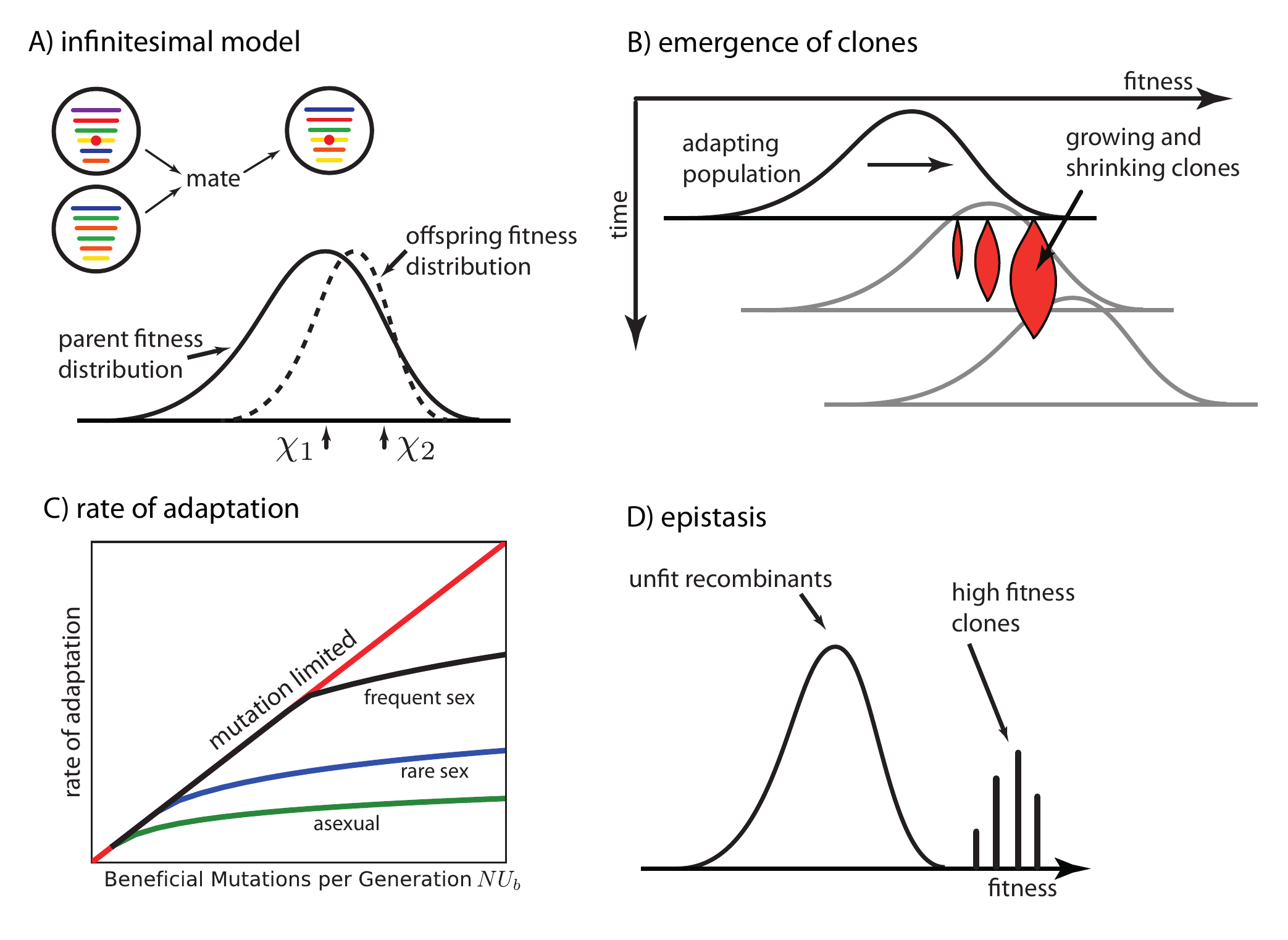}
  \caption[labelInTOC]{A facultatively sexual lifecycle is common among many
  pathogens, plants, and some groups of animals. (a) If many loci segregate
  independently, recombination can be modeled by the infinitesimal model.
  Given two parents with fitness $\xs_1$ and $\xs_2$ sampled from the
  parental distribution with variance $\sigma^2$, offspring fitness is
  symmetrically distributed around the parental mean with variance
  $\sigma^2/2$. A mutation, indicated as a red dot in the sketch, can
  thereby hop from an individual with one background fitness to a very
  different one. (b) If the outcrossing rate is lower than the fitness of
  some individuals, clones, indicated in red, can grow at rate $\xs-\ox$.
As the population adapts, the growth rate of the clones is reduced,
eventually goes negative and the clone disappears. The beneficial mutation,
however, persists on other backgrounds. In small populations, the rate of
adaptation increases linearly with the population size as sketched in panel (c).
For each outcrossing rate, there is a point beyond which interference starts to be important. 
(d) Epistasis causes condensation of the
population into a small number of very fit genotypes.
Crosses between these genotypes result in unfit individuals. In the absence of
forces that stabilize different clones, one clone will rapidly take over if
$\xs>\ox$. Scripts illustrating evolution of faculatively sexual populations
can be found in the \href{http://webdav.tuebingen.mpg.de/interference/fixation_sex.html}{online
  supplement}.}
  \label{fig:fac_sex}
\end{center}
\end{figure}

Most of our theoretical understanding of evolution in large facultatively mating
populations comes from models similar to those introduced above for asexual populations.
In addition to mutation, we have to introduce a term that describes how an
allele can move from one genetic background to another by recombination; see
\FIG{fac_sex}A.
Given the fitness values of the two parents $\xs_{1}$ and $\xs_{2}$ and
assuming many independently segregating loci, the offspring fitness $\xs$
is symmetrically distributed around the mid-parent value with
half the population variance; see illustration in
\FIG{fac_sex}A and \citep{Bulmer_1980,Turelli:1994p2652}.
To understand the process of fixation in such a population, the following is a
useful intuition: An outcrossing
event places a beneficial mutation onto a novel genotype, which is amplified by
selection into a clone whose size grows rapidly with the fitness of the
founder; see \FIG{fac_sex}B.
These clones are transient, since even an initially fit clone falls behind
the increasing mean fitness. However, large clones produce many recombinant
offspring (daughter clones), which greatly enhances the chance of fixation of
mutations they carry. Since clone size increases rapidly with founder fitness,
the fixation probability $\pfix(\xs,s)$  is still a very steep function of the
background fitness and qualitatively similar to the asexual case
(\FIG{CI_vs_MM}C).
With increasing outcrossing rate, the fitness window from which successful
clones originate becomes broader and broader.

If outcrossing rates are large enough that genotypes are disassembled by
recombination faster than selection can amplify them, $\pfix(\xs,s)$ is
essentially flat and the genetic background does not matter much. This
transition was examined by \citet{Neher:2010p30641}:
\begin{equation}
\label{eq:v_sex}
v \approx  \begin{cases}
\frac{2\ox^2 \log (NU_b)}{(\log \ox/s)^2} & \ox\ll \sqrt{NU_b s^2}\\
NU_b s^2 & \ox\geq \sqrt{NU_bs^2} \ .
\end{cases}
\end{equation}
The essence of this result is that adaptation is limited by recombination
whenever $\ox$ is smaller than the standard deviation in fitness in the absence
of interference.
In this regime, $v$ depends weakly on $N$, but increases
rapidly with $\ox$. This behavior is sketched in \FIG{fac_sex}C. Similar
results can be found in \citet{Weissman:2012p47436}.
The above analysis assumed that recombination is rare, but still frequent enough
to ensure that mutations that rise to high frequencies are essentially in linkage
equilibrium. This requires $\ox\gg s$.
\citet{Rouzine:2005p17398,Rouzine:2010p33121} studied the selection on standing
variation at intermediate and low recombination rates. 
Adaptation in presence of horizontal gene transfer was investigated by
\citet{Cohen:2005p5007}, \citet{Wylie:2010p36921}, and \citet{Neher:2010p30641}.

In contrast to asexual evolution, epistasis can dramatically affect the
evolutionary dynamics in sexual populations. Epistasis implies that the effect
of mutations depends on the state at other loci in the genome. In the absence of
sex, the only quantity that matters is the distribution of available mutations,
$\mut(s)$. The precise nature of epistasis is not crucial. 
In sexual populations, however, epistasis can affect the evolutionary dynamics
dramatically: When different individuals mix their genomes, it matters whether mutations acquired in different lineages
are compatible. Since selection favors well adapted combinations of alleles,
recombination is expected to be on average disruptive and recombinant offspring
have on average lower fitness than their parents (the so-called ``recombination
load'').
This competition between selection for good genotypes and recombination can result in a condensation of
the population into fit clones; see \FIG{fac_sex}D,
\citet{Neher:2009p22302} and \citet{neher_emergence_2013}.

\section{Selective interference in obligately sexual organisms}
Selective interference has historically received most attention in obligately
sexual organisms most relevant to crop and animal breeding.
Artificial selection has been performed by farmers and breeders for thousands of
years with remarkable success \citep{Hill:2010p48133}. Evolution experiments
with diverse species, including chicken, mice and Drosophila, have shown that
standing variation at a large number of loci responds to diverse selection
pressures
\citep{Chan:2012p48098,Burke:2010p42546,turner_population-based_2011,zhou_experimental_2011,johansson_genome-wide_2010};
see \citet{burke_how_2012} for a recent review.
In obligately sexual populations, distant loci can respond independently to
selection and remain in approximate linkage equilibrium. The
frequencies of different alleles change according to their effect on fitness
averaged over all possible fitness backgrounds in the population. Small deviations from
linkage equilibrium can be accounted for perturbatively using the so-called
Quasi-Linkage Equilibrium (QLE) approximation
\cite{Kimura:1965p3008,Barton:1991p2659,Neher:2011p45096}. 

\begin{figure}[htp]
\begin{center}
  \includegraphics[width=0.7\columnwidth]{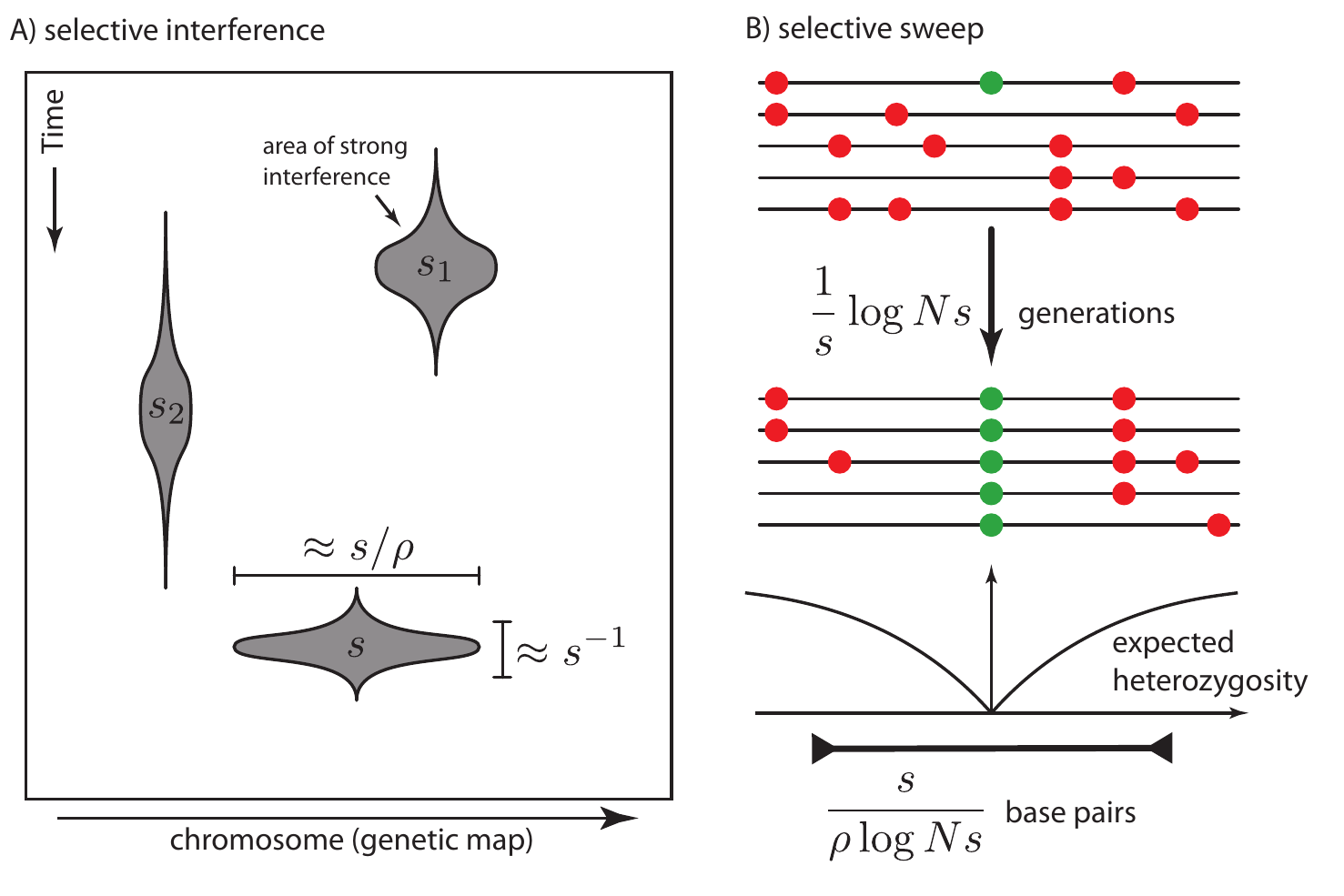}
  \caption[labelInTOC]{Interference in obligately sexual populations. Panel (a)
  sketches the interference effects of selective sweeps through time (vertical axis) and
  along the genome (horizontal axis). A sweeping mutation with selection
  coefficient $s$ interferes with other mutation in a region of width $s/\xo$
  over a time $s^{-1}$, where $\xo$ is the crossover rate per base.
  The extent of interference is sketched by grey bulges, each of which
  corresponds to a mutation that fixed.
  Interference starts to be important when the bulges overlap. Since the area of the bulges,
  roughly ``height$\times$width'', is approximately
  independent of $s$, interference depends on $\xo$ and the rate of sweeps
  rather than the effect size.
  The rate of adaptation is therefore primarily a function of the maplength $R$.
  (b) A selective sweep reduces neutral genetic variation in a region of width
  $s/(\xo \log(Ns))$. The effect of sweeps on neutral diversity is explored in
  \href{http://webdav.tuebingen.mpg.de/interference/draft.html}{online
  supplement}
  }
  \label{fig:obligate_sex}
\end{center}
\end{figure}

This approximate independence, however, does not hold for loci that are tightly
linked. \citet{Hill:1966p21029} observed that interference between linked
competing loci can slow down the response to selection -- an effect now termed
\emph{Hill-Robertson interference} \citep{Felsenstein:1974p23937}. Felsenstein
realized that interference is not restricted to competing beneficial mutations
but that linked deleterious mutations also impede fixation of beneficial
mutations (see background selection below). The term Hill-Robertson interference
is now used for any reduction in the efficacy of selection caused by linked fitness
variation. A deeper understanding of selective interference was gained in the
1990ies \citep{Barton:1994p34628,Barton:1995p3540}.
The key insight of Barton was to calculate the fate of a novel mutation
considering all possible genetic backgrounds on which it can arise and summing
over all possible trajectories it can take through the population. For a small
number of loci, the equations describing the probability of fixation can be integrated
explicitly.

Weakly-linked sweeps cause a cumulative reduction of the fixation probability at
a focal site that is roughly given by the ratio of additive variance in fitness
and the squared degree of linkage \citep{Barton:1995p3540,Santiago:1998p34629}.
\citet{Barton:1994p34628} further identified a critical rate of strong selective
sweeps that effectively prevents the fixation of mutations with an advantage smaller than $s_c$. If sweeps are
too frequent, the weakly selected mutation has little chance of spreading before
its frequency is reduced again by the next strong sweep.

At short distances, selective sweeps impede each other's fixation more strongly.
This interference is limited to a time interval of order $s^{-1}$ generations
where one of the sweeping mutations is at intermediate frequencies. During this time, a new
beneficial mutation will often fall onto the wildtype background and is lost
again if it is not rapidly recombined onto the competing sweep. The latter is
likely only if it is further than $s/\xo$ nucleotides away from the competing
sweep, where $\xo$ is the crossover rate per basepair \citep{Barton:1994p34628}.
In other words, a sweeping mutation with effect $s$ prevents other sweeps in a region of width $s/\xo$, and occupies
this chromosomal ``real estate'' for a time $s^{-1}$; see \FIG{obligate_sex}A
\citep{Weissman:2012p47436}. Hence strong sweeps briefly interfere with other
sweeps in a large region, while weak sweeps affect a narrow region for a longer
time. The amount of interference is therefore roughly independent of the
strength of the sweeps, and the total number of sweeps per unit time is limited
by the map-length $R = \int  \xo(y)\, dy$, where the integral is over the entire
genome and $\rho(y)$ is the local crossover rate. Larger populations can squeeze
slightly more sweeps into $R$ \citep{Weissman:2012p47436}. In most obligately
sexual organisms, sweeps rarely cover more than a few percent of
the total map length such that recombination is not limiting adaptation unless
sweeps cluster in certain regions \citep{Sella:2009p26729}. However, as I will
discuss below, even rare selective sweeps have dramatic effects on neutral
diversity.

\section{Genetic diversity, draft, and coalescence}
Interference between selected mutations reduces the fixation probability of
beneficial mutations, slows adaptation, and weakens purifying selection. These
effects are very important, but hard to observe since significant adaptation
often takes longer than our window of observation. Typically, data consists of a
sample of sequences from a population. These sequences differ by single
nucleotide polymorphisms, insertions, or deletions, and we rarely know the
effect of these differences on the organism's fitness.

From a sequence sample of this sort, the genealogy of the population is
reconstructed and compared to models of evolution -- in most cases a neutral
model governed by Kingman's coalescent \citep{Kingman:1982p28911}. From this
comparison we hope to learn about evolutionary processes. However, linked
selection, be it in asexual organisms, facultatively sexuals, or obligately
sexuals, has dramatic effects on the genealogies. Substantial effects on neutral
diversity are observed at rates of sweeps that do not yet cause strong
interference between selected loci for the simple reason that neutral alleles
segregate for longer times \citep{Weissman:2012p47436}.

\subsection{Genetic draft in obligately sexual populations}
\begin{figure}[htp]
\begin{center}
  \includegraphics[width=0.6\columnwidth]{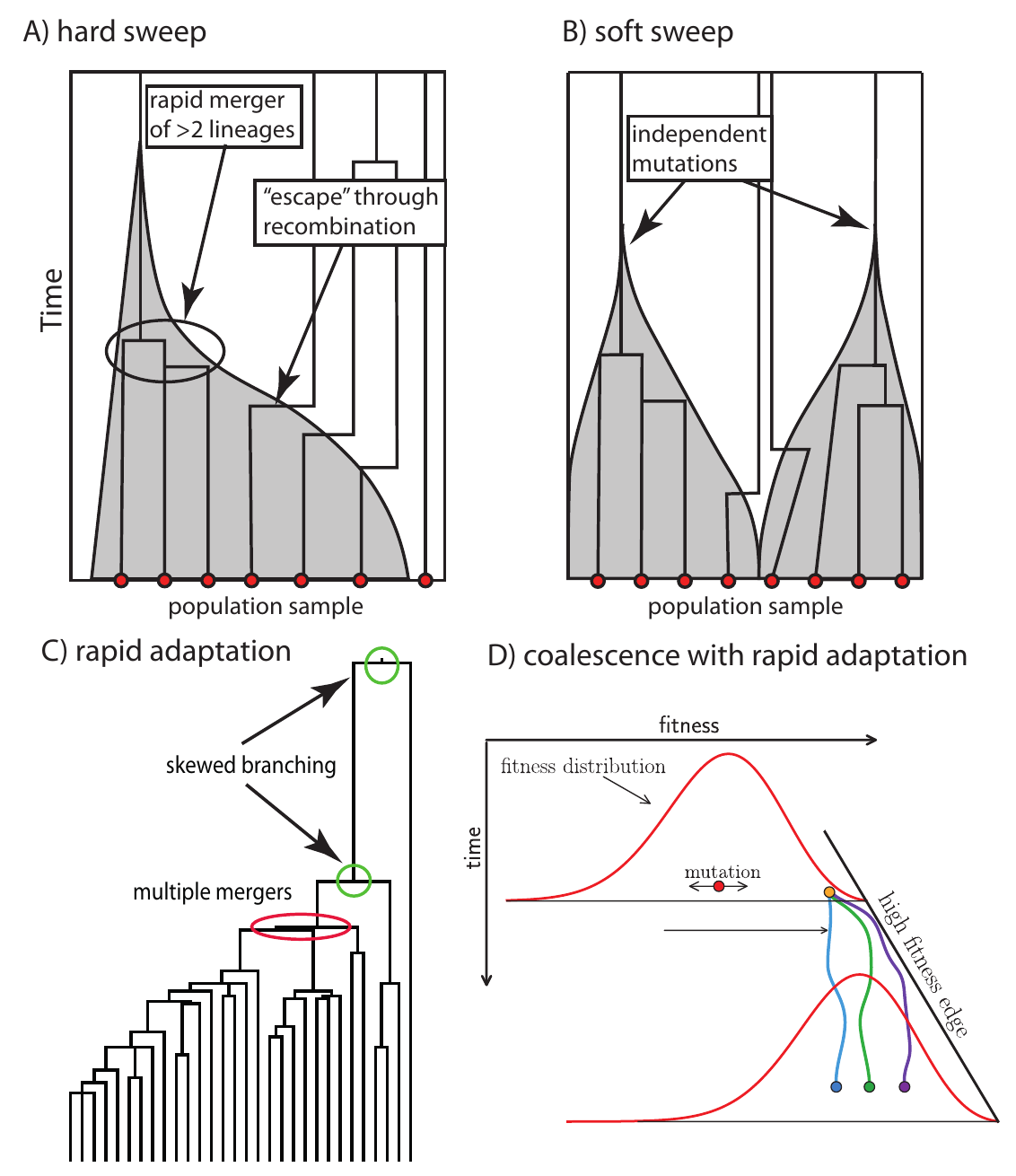}
  \caption[labelInTOC]{Coalescence driven by selection. (a) A selective sweep
  (grey region) causes rapid coalescence of lineages at a nearby locus. Each
  sweep causes a fraction of lineages to merge, while the remainder recombines onto an
  ancestral background. (b)
  Soft sweeps refer to a scenario where single mutations arise multiple times  
  independently in response to environmental change. This is expected as soon
  as the product of $N$ and the per site mutation rate exceeds one and can
  result in multiple bursts of coalescence almost at the same time.
  (c) A genealogical tree drawn from a simulation of a model of rapidly adapting
  asexual organisms. Coalescence often occurs in bursts. Furthermore, branching
  is often uneven. At many branchings in this ``ladderized'' tree, most
  individuals descend from the left branch.
  Those are well known features of multiple merger coalescence processes such as the Bolthausen-Sznitman coalescent.
  (d) Coalescence and fitness classes. Most population samples consists of
  individual from the center of the fitness distribution, while their
  distant ancestors were among the fittest. In large populations, most
  coalescence happens in the high fitness nose and the time until ancestral
  lineages ``arrive" in the  nose corresponds to long terminal branches
  (compare panel c). How genealogies depend
  on selection can be studied using simulations, see
  \href{http://webdav.tuebingen.mpg.de/interference/coalescence.html}{online
  supplement}.
 }
  \label{fig:coalescence}
\end{center}
\end{figure}

Selective sweeps have strong effects on linked neutral diversity and genealogies
\citep{Smith:1974p34217,Kaplan:1989p34931,Stephan:1992p35204,Wiehe:1993p37333,Barton:1998p28270,Barton:2004p34826}.
A sweeping mutation takes about $t_{sw}\approx s^{-1}\log Ns$ generations to
rise to high frequency. Linked neutral variation is preserved only when
substantial recombination happens during this time. Given a crossover rate
$\xo$ per base, recombination will separate the sweep from a locus at distance
$l$ with probability $r = \xo l$ per generation (assuming $r\ll 1$). 
Hence a sweep leaves a dip of width $l=(\xo t_{sw})^{-1}\approx s/(\rho\log Ns)$
in the neutral diversity (see \FIG{obligate_sex}B).
Within this region, selection causes massive and rapid coalescence and only a
fraction of the lineages continue into the ancestral population (see
\FIG{coalescence}A).
This effect has been further investigated by \citet{Durrett:2005p40919}, who
showed that the effect of recurrent selective sweeps is well approximated by a
coalescent process that allows for multiple mergers: each sweep forces the
almost simultaneous coalescence of a large number of lineages (a fraction
$e^{-\ox t_{sw}}$). Similar arguments had been made previously by
\citet{Gillespie:2000p28513}, who called the stochastic force responsible for
coalescence {\it genetic draft}.
\citet{coop_patterns_2012} extended the analysis of Durret and Schweinsberg
partial sweeps that could be common in structured populations, with 
over-dominance, or frequency dependent selection.

The rapid coalescence of multiple lineages is unexpected in the standard neutral
coalescent (a merger of $p$ lineages occurs with probability $\propto N^{-p}$).
In coalescence induced by a selective sweep, however, multiple mergers are common
and dramatically change the statistical properties of genealogies. 
A burst of coalescence corresponds to a portion of the tree with almost
star-like shape \citep{Slatkin:1991p43283}. Alleles that arose before the burst
are common, those after the burst rare. This causes a relative increase of rare
alleles, as well as alleles very close to fixation
\citep{Braverman:1995p34932,Fay:2000p35077,Gillespie:2000p28513}. 

The degree to which linked selective sweeps reduce genetic diversity depends
primarily on the rate of sweeps per map length \citep{Weissman:2012p47436}. In
accord with this expectation, it is found that diversity increases with
recombination rate and decreases with the density of functional sites
\citep{Begun:2007p9546,shapiro_adaptive_2007}.
In addition to occasional selective sweeps, genetic diversity and the degree of
adaptation can be strongly affected by a large number of weakly selected sites,
e.g.~weakly deleterious mutations, that generate a broad fitness distribution
\citep{McVean:2000p19278}.

\subsection{Soft sweeps}
Soft sweeps refer to events when a
selective sweep originates from multiple genomic backgrounds
\citep{Hermisson:2005p1649,Pennings:2006p35571}, either because the favored
allele arose independently multiple times or because it has been segregating
for a long time prior to a environmental change.
Soft sweeps have recently been observed in pesticide resistance of Drosophila
\citep{Karasov:2010p35377} and are a common phenomenon in viruses with high
mutation rates.

A genealogy of individuals sampled after a soft sweep is illustrated in
\FIG{coalescence}B. The majority of the individuals trace back to
one of two or more ancestral haplotypes on which the selected mutation arose.
Hence coalescence is again dominated by multiple merger events, except that
several of those events happen almost simultaneously. This type of coalescent process has
been described in \citet{Schweinsberg:2000p40940}. 

Despite dramatic effects on genealogies, soft sweeps can be difficult to
detect by standard methods that scan for selective sweeps. Those methods use
local reductions in genetic diversity, which can be modest if the population
traces back to several ancestral haplotypes. The number of ancestral haplotypes
in a sample after a soft sweep depends on the product of $N$, the per-site
mutations rate $\mu$, and selection against the allele before the sweep
\citep{Pennings:2006p35571}. To detect soft sweeps, methods are required that
explicitly search for signatures of rapid coalescence into several lineages
in linkage disequilibrium or haplotype patterns
\citep{Pennings:2006p35571,Messer:2012p47356}.

\subsection{The Bolthausen-Sznitman coalescent and rapidly adapting populations}
Individual selective sweeps have an intuitive effect on genetic diversity, but
what do genealogies look like when many mutations are competing in asexual or
facultatively sexual populations? It has recently been argued that the
genealogies of populations in many models of rapid adaptation are well described
by coalescent processes with multiple mergers
\citep{Pitman:1999p41045,Berestycki:2009p45808}. 
This was first discovered by \citet{Brunet:2007p18866},
who studied a model where a population expands its range. The genealogies of
individuals at the front are described by the Bolthausen-Sznitman coalescent, a
special case of coalescent processes with multiple mergers. Recently, it has
been shown that a similar coalescent process emerges in models of adaptation in
panmictic populations \citep{neher_genealogies_2013,desai_genetic_2012}.

\FIG{coalescence}C shows a tree sampled from a model of a rapidly adapting
population.
A typical sample from a rapidly adapting population will consist of individuals
from the center of the fitness distribution. Their ancestors tend to be among
the fittest in the population \citep{Rouzine:2007p17401,Hermisson:2002p47231}.
Substantial coalescence happens only once the ancestral lineages have reached
the high fitness tip, resulting in long terminal branches of the trees. Once in
the tip, coalescence is driven by the competition of lineages against each other
and happens in bursts whenever one lineage gets ahead of everybody else. These
bursts correspond to the event that a large fraction of the population descends
from one particular individual. These coalescent events have
approximately the same statistics as neutral coalescent processes with very
broad but non-heritable offspring distributions
\citep{Eldon:2006p36003,Schweinsberg:2003p40932,Der:2011p44500}.

In the case of rapidly adapting asexual populations, the effective distribution
of the number $n$ of offspring is given by $P(n)\sim n^{-2}$ which gives rise to
the Bolthausen-Sznitman coalescent.
This type of distribution seems to be universal to populations in which
individual lineages are amplified while they diversify and is found in
facultatively sexual populations \citep{Neher:2011p42539}, asexual populations
adapting by small steps, as well as populations in a dynamic balance between
deleterious and beneficial mutations.
Asymptotic features of the site frequency spectrum can be derived analytically
\citep{Berestycki:2009p45808,neher_genealogies_2013,desai_genetic_2012}.
One finds that the frequency spectrum diverges as $f(\nu) \sim \nu^{-2}$ at low
frequencies corresponding to many singletons. Furthermore,
neutral alleles close to fixation are common with $f(\nu)$ diverging again as $\nu\to
1$. This relative excess of rare and very common alleles is a consequence
multiple mergers which produce star-like sub-trees and the very asymmetric
branching at nodes deep in the tree (compare \FIG{coalescence}C).

The time scale of coalescence, and with it
the level of genetic diversity, is mostly determined by the strength
of selection and only weakly increases with population size. Essentially, the
average time to a common ancestor of two randomly chosen individuals is given by
the time it takes until the fittest individuals dominate the
population. In most models, this time depends only logarithmically on the
population size $N$.

\subsection{Background selection and genetic diversity}
Background selection refers to the effect of purifying selection on linked loci,
which is particularly important if linked regions are long. If deleterious
mutations incur a fitness decrement of $s$ and arise with genome wide rate
$U_d$, a sufficiently large population settles in a state where the number of mutations
in individuals follows a Poisson distribution with mean $\lambda =
U_d/s$ \citep{Haigh:1978p37141}. Individuals loaded with many mutations are
selected against, but continually produced by de novo mutations. All
individuals in the population ultimately descent from individuals carrying least
deleterious mutations. Within this model, the least loaded class has size
$N\exp(-U_d/s)$ and coalescence in this class is accelerated by $\exp(U_d/s)$
compared to a neutrally evolving population of size $N$
\citep{Charlesworth:1993p36005}. For large ratios $U_d/s$, the Poisson
distribution of background fitness spans a large number of fitness
classes and this heterogeneity substantially reduces the efficacy of selection
\citep{McVean:2000p19278}.

The effect of background selection is best appreciated in a genealogical
picture. Genetic backgrounds sampled from the population tend to come from the center of the distribution.
Since the deleterious mutations they carry were accumulated in the recent past,
lineages ``shed'' mutations as we trace them back in time until they arrive in
the mutation free class akin to \FIG{coalescence}D.
This resulting genealogical process, a fitness class coalescent, has been
described in \citet{Walczak:2011p45228}.
A recent study on the genetic diversity of whale lice \citep{Seger:2010p39561}
suggests that purifying selection and frequent deleterious mutations can
severely distort genealogies. \citet{Ofallon:2010p43263} present methods for the
analysis of sequence samples under purifying selection.

The fitness class coalescent is appropriate as long as Muller's ratchet does not
yet click. More generally, fixation of deleterious mutations, adaptation,
and environmental change will balance approximately. It has been shown that a
small fraction of beneficial mutations can be sufficient to halt Muller's
ratchet \citep{Goyal:2012p47382}. In this dynamic balance between
frequent deleterious and rare beneficial mutations, the genealogies tend to be
similar to genealogies under rapid adaptation discussed above.

\section{Conclusions and future directions}
Contradicting neutral theory, genetic diversity correlates only weakly with
population size \citep{leffler_revisiting_2012}, suggesting that linked
selection or genetic draft are more important than conventional genetic drift.
Draft is most severe in asexual populations, for which models predict that
the fitness differences rather than the population size determine the level of
neutral diversity.
As outcrossing becomes more frequent, the strength of draft decreases and
diversity increases. With increasing coalescence times, selection becomes more efficient
as there is more time to differentiate deleterious from beneficial alleles. In
obligately sexual populations, most interference is restricted to tightly linked
loci and the number of sweeps per map length and generation determines genetic
diversity.

Since interference slows adaptation, one expects that adaptation can select for
higher recombination rates \citep{Charlesworth:1993p25074}.
Indeed, positive selection results in indirect selection on recombination
modifiers
\citep{Barton:1995p2811,Otto:1997p1074,Barton:2005p982,Hartfield:2010p34731}.
Changing frequencies of outcrossing have been observed in evolution experiments
\citep{Becks:2010p43116}. However, the evolution of recombination and
outcrossing rates in rapidly adapting populations remains poorly understood,
both theoretically and empirically.

The traveling wave models discussed above assume a large number of polymorphisms
with similar effects on fitness and a smooth fitness distribution,
which are drastic idealizations.
More typically, one finds a handful of polymorphisms with a distribution of
effects \citep{Barrick:2009p39413,Lang:2011p41792,Strelkowa:2012p48466}.
Simulations indicate, however, that statistical properties of genealogies are
rather robust regarding model assumptions as long as draft dominates over drift
\citep{neher_genealogies_2013}.
Appropriate genealogical models are prerequisite for demographic
inference. If, for example, a neutral coalescent model is used to infer the
population size history of a rapidly adapting population, one would conclude
that the population has been expanding. Incidentally, this is inferred in most
cases.
Some progress towards incorporating the effect of purifying selection into
estimates from reconstructed genealogies has been made recently
\citep{Ofallon:2011p43932,nicolaisen_distortions_2012}.
Alternative genealogical models accounting for selection should be included into
popular analysis programs such as BEAST \citep{Drummond:2007p30127}.

It is still common to assign an ``effective'' size, $N_e$, to various
populations.
In most cases, $N_e$ is a proxy for genetic diversity, which depends on the time
to the most recent common ancestor. With the realization that coalescence times
depend on linked selection and genetic draft, rather than the population size
and genetic drift, the term should be avoided and replaced by $T_c$, the time
scale of coalescence. Defining $N_e$ suggests that the neutral model is valid as
long as $N_e$ is used instead of $N$. We have seen multiple times that drift and
draft are of rather different natures and that this difference cannot be
captured by a simple rescaling. Each quantity then requires its own private $N_e$,
rendering the concept essentially useless. Some quantities like site
frequency spectra are qualitatively different and no $N_e$ maps them to a
neutral model. The (census) population size is nevertheless important in
discovering beneficial mutations. For this reason, large populations are
expect to respond more quickly to environmental change as we are painfully aware
in the case of antibiotic resistance of pathogens. Large populations might
therefore track phenotypic optima more closely resulting in beneficial
mutations with smaller effect, which in turn might explain their greater
diversity.

The majority of models discussed assume a time invariant fitness landscape. This
assumption reflects our ignorance regarding the degree and timescale of
environmental fluctuations  (for work on selection in time-dependent fitness
landscapes, see \citet{Mustonen:2009p25388}).
Time-variable selection pressures, combined with spatial variation, could
potentially have strong effects. Similarly, frequency-dependent selection and
more generally the interaction of evolution with ecology are important avenues
for future work. The challenge consists of choosing useful models that are
tractable, appropriate, and predictive.

\acknowledgements{
I would like to thank Boris Shraiman, Sally Otto, Frank Chan, Philipp
Messer, Aleksandra Walczak and Michael Desai for stimulating discussions and
helpful comments on the manuscript. This work is supported by the ERC starting grant HIVEVO 260686.}

\appendix
\section{Glossary}
\begin{itemize}
  \item {\sc Genetic drift}: stochastic changes in allele frequencies due
  to non-heritable variation in offspring number.
  \item {\sc Purifying selection}: selection against deleterious mutations.
  \item {\sc Positive selection} : selection for novel beneficial mutations.
  \item {\sc Genetic draft}: changes in allele frequencies due to (partly)
  heritable random associations with genetic backgrounds.
  \item {\sc Hitchhiking}: rapid rise in frequency through an association
  with a very fit background.
  \item {\sc Selective interference}: reduction of fixation probability
  through competition with other beneficial alleles.
  \item {\sc Clonal interference}: competition between well adapted asexual
  subpopulations from which only one subpopulation emerges as winner.
  \item {\sc Branching process}: stochastic model of reproducing and dying
  individuals without a constraint on the overall population size.
  \item {\sc Epistasis}: background dependence of the effect of mutations.
  Epistasis can result in rugged fitness landscapes.
  \item {\sc Kingman coalescent}: basic coalescence process where random
  pairs of individuals merge.
  \item {\sc Multiple merger coalescent}: coalescent process with simultaneous
  merging of more than 2 lineages.
  \item {\sc Bolthausen-Sznitman Coalescent (BSC)}: special multiple
  merger coalescent which approximates genealogies in many models of adaptation.
\end{itemize}
\newpage

\end{document}